\documentclass[preprint,12pt]{elsarticle}
\usepackage{amssymb}

\begin{document}

\begin{frontmatter}

\title{On Quantum Turing Machine Halting Deterministically}

\author{Min Liang}
\author{Li Yang\corref{1}}\ead{yangli@iie.ac.cn}
\cortext[1]{Corresponding author.}
\address{Institute of Information Engineering, Chinese Academy of Sciences, Beijing, China}
\address{Graduate University of Chinese Academy of Sciences, Beijing, China}

\begin{abstract}
We define a subclass of quantum Turing machine (QTM) named SR-QTM, which halts deterministically and has deterministic tape head position. A quantum state transition diagram (QSTD) is proposed to describe SR-QTM. With the help of QSTD, we construct a SR-QTM which is universal for all near-trivial transformations. This means there exists a QTM which is universal for the above subclass. Finally we prove that SR-QTM is computational equivalent with ordinary QTM in the bounded error setting. It can be seen that, because SR-QTM has the same time steps for different branches of computation, the halting scheme problem will not exist when considering SR-QTM as a model of quantum computing.
\end{abstract}


\end{frontmatter}


\section{Introduction}
\label{}
Deutsch \cite{deutsch85} formulates the model of quantum computer which is referred to as quantum Turing machine (QTM). Then it is pointed out that there exists a halting scheme problem in case that a QTM takes different time steps for different branches of computation \cite{myers97}. This problem is avoided by Bernstein and Vazirani \cite{bernstein97} who considers QTM that has the same time steps for all different branches. The universal quantum Turing machine is constructed in this context.

Ozawa \cite{ozawa98} proves that the halting scheme is equivalent to the quantum nondemolition (QND) monitoring of the output observable. Linden and Popescu \cite{linden98} argue that the halting scheme proposed by Ozawa \cite{ozawa98} is not consistent with unitarity of the evolution operator and the halting scheme problem is still unsolved. Then Ozawa proposes QTM with well-behaved halt flags, and arbitrary QTM can be simulated by this kind of QTM \cite{ozawa02}. Moreover, this kind of QTMs obeys a halting protocol, which requires one measurement of the halt flag after one time step. He shows that the halting protocol does not affect the computing result of QTM. That means the probability distribution of the output is not affected by monitoring of the halt flag. Thus the halt scheme problem is solved though the QTM may halt probabilistically.

Miyadera and Ohya \cite{miyadera05} prove it is impossible to efficiently distinguish between those QTMs which halts deterministically and those which halts probabilistically. There are also some discussions about halting of universal QTM \cite{kieu01,shi02,fouche07}.

In this paper, we consider those QTMs which halts deterministically. Firstly, we define a subclass of QTM (called SR-QTM) which has the same time steps for all branches of computation and has deterministic tape head position. Then we propose a new way named quantum state transition diagram (QSTD) to describe SR-QTM. It is proved that any quantum circuit consisting of CNOT and single-qubit gates can be efficiently simulated by SR-QTM. Then, based on universal quantum circuit of near-trivial transformations \cite{liang11}, we present a SR-QTM which is universal for all near-trivial transformations \cite{bernstein97}, so we have shown that there exists a QTM which is universal for a subclass of QTM. Finally, the computational power of SR-QTM is analyzed and we prove the computational equivalence (in the bounded error setting) between QTM and SR-QTM. Since SR-QTM halts deterministically, the halting scheme problem dissolve in this sense.


\section{Stationary rotational QTM}
\label{}
QTM is defined by a triplet ($\Sigma$, $Q$, $\delta$) \cite{bernstein97}, where $\Sigma$ is a finite alphabet with an identified blank symbol $\#$, $Q$ is a finite set of states with an identified initial state $q_0$, and final state $q_f$ ($q_f\neq q_0$), and  $\delta$, the quantum transition rules, is a function
\begin{displaymath}
\delta : Q\times\Sigma\times\Sigma\times Q\times \{\leftarrow,\rightarrow\}\mapsto \widetilde{\mathbb{C}},
\end{displaymath}
where '$\leftarrow$' and '$\rightarrow$' indicate the direction in which the tape head move, and $\widetilde{\mathbb{C}}$=\{$\alpha\in \mathbb{C}$ $|$ there is a deterministic algorithm that computes the real and imaginary parts of $\alpha$ to within $2^{-n}$ in time polynomial in $n$\}. In the process of quantum computing, the state may be transformed into certain superposition of states in $Q$. QTM measures its state after each step of computing, and the state collapses to one component state. Thus the QTM collapses to certain computational branch and continues its next computation according to the transition function.

{\bf Definition 1}. A QTM is called stationary QTM, if it takes the same time steps for all branches of computation, and its tape head always go back to the start cell when the computation terminates.

{\bf Definition 2}. A QTM is called unidirectional QTM \cite{bernstein97}, if $d_1=d_2$ whenever $\delta(p_1, \sigma_1, \tau_1, q, d_1)\neq 0$ and $\delta(p_2, \sigma_2, \tau_2, q, d_2)\neq 0$.

For a unidirectional QTM, the transition $\delta(p, \sigma, \tau, q, d)=\alpha(\alpha\in\widetilde{\mathbb{C}})$ is written as $\delta(p, \sigma, \tau, q, d(q))=\alpha$ because the transition direction $d$ is decided only by the state $q$.

Next we define rotational QTM and SR-QTM as follows:

{\bf Definition 3}. A QTM is called rotational QTM, if it is unidirectional and satisfies the condition: $q_1=q_2$ whenever $\delta(p, \sigma, \tau_1, q_1, d_1)\neq 0$ and $\delta(p, \sigma, \tau_2, q_2, d_2)\neq 0$.

From our definition of rotational QTM, it can be included that the state to be entered is only related to current state and the content in the tape head. Thus, the transition function of rotational QTM such as $\delta(p, \sigma, \tau, q, d)=\alpha$ can be written as $\delta(p, \sigma, \tau, q(p,\sigma), d(q))=\alpha$.

We say a QTM has deterministic tape head position \cite{watrous95}, if this QTM is observed at any time during its computation, the probability that the tape head will be observed in any given location will be either 0 or 1.

{\bf Definition 4}. A QTM is called stationary rotational QTM (SR-QTM) if it has deterministic tape head position, and satisfies the conditions in both Definition 1 and Definition 3.

\section{Quantum state transition diagram}

In order to further understand SR-QTM, we present a new method called quantum state transition diagram (QSTD) to describe SR-QTM. The QSTD for QCA(quantum cellular automata) is proposed by Hook and Lee \cite{hook10}, but it is different from ours.

By the definition of SR-QTM, the state $q$ in the transition $\delta(p, \sigma, \tau, q, d)=\alpha$ is only related with $p$ and $\sigma$, and the transition direction $d$ only depends on $q$, so the transition can be written as $\delta(p, \sigma, \tau, q(p,\sigma), d(q))=\alpha$.

\begin{figure}[htb!]
\begin{center}
\includegraphics[scale=0.9]{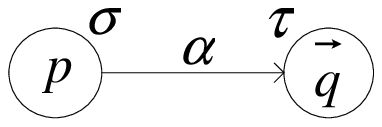}
\end{center}
\vspace{-5mm}
\caption{\label{fig1}The description of each transition rule $\delta(p, \sigma, \tau, q(p,\sigma), d(q))=\alpha$.}
\end{figure}

In QSTD (for instance, Fig.\ref{fig1}), there are two kinds of elements: circle with an alphabet in it, and the line connecting two circles. In the circle, the alphabet represents the state, and the arrow over the alphabet indicates the direction in which the tape head move (the alphabet without arrow is the start state). In each line connecting two circles, there is only one arrow which indicates the state which the current state is transferring to. The alphabet on the tail/head of the line represents the content to be read/write. In the middle of line, a number is marked to represent the transition amplitude.

For a given SR-QTM, we can draw a QSTD to describe it according to the following two rules:

(1) For each transition rule of SR-QTM such as $\delta(p, \sigma, \tau, q(p,\sigma), d(q))=\alpha$, where $d(q)\in\{\leftarrow,\rightarrow\}$, the QSTD can be represented in Fig.\ref{fig1} (suppose $d(q)=\rightarrow$). If the transition amplitude $\alpha$ is equal to 1, the mark $\alpha$ in the QSTD could be omitted.

\begin{figure}[htbp]
\begin{center}
\includegraphics[scale=0.9]{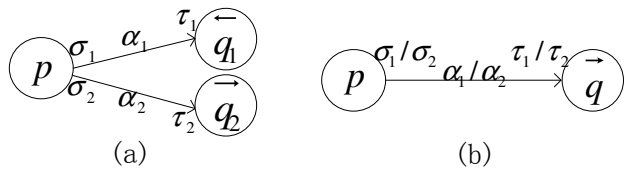}
\vspace{-3mm}
\caption{\label{fig2}The description of transitions rules $\delta(p, \sigma_i, \tau_i, q_i, d(q_i))=\alpha_i$, where $q_i=q(p,\sigma_i),i=1,2$. The QSTD (b) is a simplified version of the QSTD (a) in the condition that $q_1=q_2=q$.}
\end{center}
\end{figure}

(2) For any two transition rules such as $\delta(p, \sigma_i, \tau_i, q_i, d(q_i))=\alpha_i$, where $q_i=q(p,\sigma_i),i=1,2$, the QSTD is shown in the left diagram of Fig.\ref{fig2} (suppose $d(q_1)=\leftarrow$ and $d(q_2)=\rightarrow$). Moreover, if $q_1=q_2=q$, then $d(q_1)=d(q_2)=d(q)$. Suppose $d(q)=\rightarrow$, the left QSTD can be simplified as the right one in Fig.\ref{fig2}. Note that in this case we can infer from the unitarity of QTM that $\tau_1\neq\tau_2$, however $\sigma_1$ may be equal to $\sigma_2$ or not.

Finally, an example is presented for better understanding QSTD. Suppose the tape of QTM is one-way infinite and its cells are numbered with $0, 1, \cdots, n, \cdots$. The start cell (cell $0$) is always initiated with $|\#\rangle$.

We construct a SR-QTM which performs Hadamard transformation on the cell 2. The SR-QTM is ($\Sigma$, $Q$, $\delta$), where $\Sigma=\{\#,0,1\}$, $Q=\{q_0,q_1,q_2,q_3,q_f\}$, and $\delta$ is as follows.
\begin{eqnarray*}
&\delta(q_0,\#,\#,q_1,\rightarrow)=1,\\
&\delta(q_1,0,0,q_2,\rightarrow)=1,                 &\delta(q_1,1,1,q_2,\rightarrow)=1, \\
&\delta(q_2,0,0,q_3,\leftarrow)=\frac{1}{\sqrt{2}}, &\delta(q_2,0,1,q_3,\leftarrow)=\frac{1}{\sqrt{2}},\\
&\delta(q_2,1,0,q_3,\leftarrow)=\frac{1}{\sqrt{2}}, &\delta(q_2,1,1,q_3,\leftarrow)=-\frac{1}{\sqrt{2}},\\
&\delta(q_3,0,0,q_f,\leftarrow)=1,                  &\delta(q_3,1,1,q_f,\leftarrow)=1.
\end{eqnarray*}

We can verify this SR-QTM implements Hadamard transformation on cell $2$. Suppose the first three cells are initiated with arbitrary state $|\#\rangle(a|00\rangle+b|01\rangle+c|10\rangle+d|11\rangle)$, where $|a|^2+|b|^2+|c|^2+|d|^2=1$. Initially, SR-QTM is in the start state $q_0$ and the tape head stays on cell $0$, so the initial configuration of the SR-QTM is $|q_0\rangle|\#\rangle(a|00\rangle+b|01\rangle+c|10\rangle+d|11\rangle)|0\rangle$, where the last quantum state $|0\rangle$ represents the position of the tape head. The computing process can be represented as follows.
\begin{eqnarray*}
&&|q_0\rangle|\#\rangle(a|00\rangle+b|01\rangle+c|10\rangle+d|11\rangle)|0\rangle\\
&\stackrel{1}{\longrightarrow}&|q_1\rangle|\#\rangle(a|00\rangle+b|01\rangle+c|10\rangle+d|11\rangle)|1\rangle\\
&\stackrel{2}{\longrightarrow}&|q_2\rangle|\#\rangle(a|00\rangle+b|01\rangle)|2\rangle+|q_3\rangle|\#\rangle(c|10\rangle+d|11\rangle)|2\rangle\\
&=&a|q_2\rangle|\#\rangle|00\rangle|2\rangle+b|q_2\rangle|\#\rangle|01\rangle)|2\rangle+c|q_2\rangle|\#\rangle|10\rangle|2\rangle+d|q_2\rangle|\#\rangle|11\rangle|2\rangle\\
&\stackrel{3}{\longrightarrow}&\frac{a}{\sqrt{2}}|q_3\rangle|\#\rangle|00\rangle|1\rangle+\frac{a}{\sqrt{2}}|q_3\rangle|\#\rangle|01\rangle|1\rangle+\frac{b}{\sqrt{2}}|q_3\rangle|\#\rangle|00\rangle|1\rangle-\frac{b}{\sqrt{2}}|q_3\rangle|\#\rangle|01\rangle|1\rangle\\ &&+\frac{c}{\sqrt{2}}|q_3\rangle|\#\rangle|10\rangle|1\rangle+\frac{c}{\sqrt{2}}|q_3\rangle|\#\rangle|11\rangle|1\rangle+\frac{d}{\sqrt{2}}|q_3\rangle|\#\rangle|10\rangle|1\rangle-\frac{d}{\sqrt{2}}|q_3\rangle|\#\rangle|11\rangle|1\rangle\\
&=&|q_3\rangle|\#\rangle(a|0\rangle\frac{|0\rangle+|1\rangle}{\sqrt{2}})|1\rangle+|q_3\rangle|\#\rangle(b|0\rangle\frac{|0\rangle-|1\rangle}{\sqrt{2}})|1\rangle\\ &&+|q_3\rangle|\#\rangle(c|1\rangle\frac{|0\rangle+|1\rangle}{\sqrt{2}})|1\rangle+|q_3\rangle|\#\rangle(d|1\rangle\frac{|0\rangle-|1\rangle}{\sqrt{2}})|1\rangle\\
&\stackrel{4}{\longrightarrow}&|q_f\rangle|\#\rangle\left(a|0\rangle\frac{|0\rangle+|1\rangle}{\sqrt{2}}+b|0\rangle\frac{|0\rangle-|1\rangle}{\sqrt{2}}+c|1\rangle\frac{|0\rangle+|1\rangle}{\sqrt{2}}+d|1\rangle\frac{|0\rangle-|1\rangle}{\sqrt{2}}\right)|0\rangle.
\end{eqnarray*}

\begin{figure}[htb!]
\begin{center}
\includegraphics[scale=0.9]{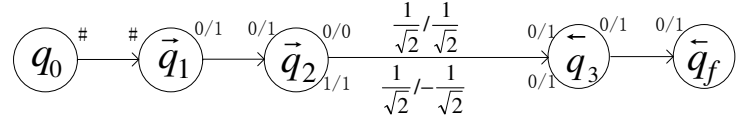}
\end{center}
\caption{\label{fig6}The QSTD of SR-QTM which carries out Hadamard transformation on cell $2$.}
\end{figure}

Then, according to the two rules of QSTD, we can draw a QSTD in Fig.\ref{fig6} to describe this SR-QTM.

\section{SR-QTM simulating quantum circuit}
\label{section3}
SR-QTM can be used to efficiently simulate quantum gates such as CNOT gate, Toffoli gate and the single-qubit rotations in $\{R_z(\pm\frac{\pi}{2^j}), R_y(\pm\frac{\pi}{2^j}),j\in{\mathbb N}\}$. In this section, we will construct some SR-QTMs simulating these quantum gates, and these SR-QTMs will be described using QSTD.

\begin{figure}[htb!]
\begin{center}
\includegraphics[scale=0.8]{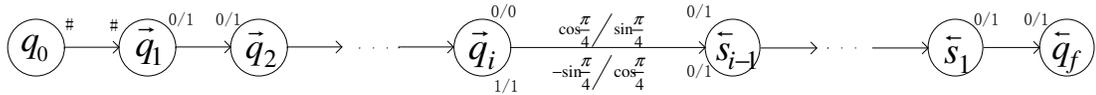}
\end{center}
\caption{\label{fig3}The QSTD of SR-QTM which carries out the quantum operation $R_y(\frac{\pi}{2})$ on cell $i$.}
\end{figure}

Firstly, we show how to construct SR-QTM which simulates certain single quit operation $U\in\{R_z(\pm\frac{\pi}{2^j}), R_y(\pm\frac{\pi}{2^j}), j\in{\mathbb N}\}$. For example, we construct a SR-QTM which performs $R_y(\frac{\pi}{2})$ on the $i$th qubit (The $i$th qubit is stored in cell $i$ of the tape). The QSTD in Fig.\ref{fig3} describes the SR-QTM ($\Sigma$, $Q$, $\delta$) which carries out the quantum operation $R_y(\frac{\pi}{2})$ on cell $i$. The set of states is $Q=\{q_0,q_1,\cdots,q_i,s_1,\cdots,s_{i-1},q_f\}$, where $q_0$ is initial state. The alphabet is $\Sigma$=\{\#,0,1\}. The tape head initially stay in the start cell, and finally go back to the start cell after carrying out the quantum operation $R_y(\frac{\pi}{2})$ on cell $i$.

\begin{figure}[htb!]
\begin{center}
\includegraphics[scale=0.95]{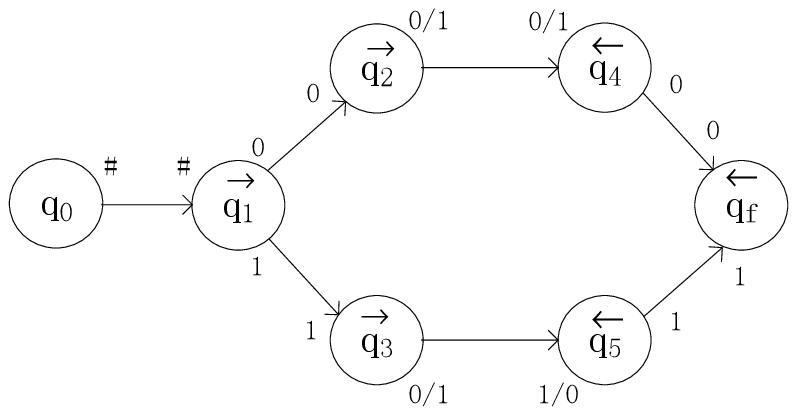}
\end{center}
\vspace{-5mm}
\caption{\label{fig4}The QSTD of SR-QTM which carries out CNOT operation between cell $1$ and cell $2$.}
\end{figure}

Then, we construct SR-QTM simulating CNOT and Toffoli gates. Suppose the blank symbol $\#$, control qubit and target qubit are stored in cell $0$, cell $1$ and cell $2$, respectively. The SR-QTM performing CNOT operation between cell $1$ and cell $2$ is described in Fig.\ref{fig4}. The set of states is $Q=\{q_0, q_1, q_2, q_3, q_4, q_5, q_f\}$, where $q_0$ is initial state. The tape head initially stay in cell $0$, and finally go back to cell $0$ after carrying out the CNOT operation between cell $1$ and cell $2$. It should be noticed that, similar to the construction in Fig.\ref{fig3}, the CNOT operation between any two cells can be efficiently simulated by a SR-QTM.


A simple verification is given to prove that the SR-QTM in Fig.\ref{fig4} can realize the CNOT operation.

Suppose the initial configuration of the SR-QTM is $|q_0\rangle|\#\rangle(a|00\rangle+b|01\rangle+c|10\rangle+d|11\rangle)|0\rangle$, where $|q_0\rangle$ is initial state of SR-QTM, $|\#\rangle(a|00\rangle+b|01\rangle+c|10\rangle+d|11\rangle)|0\rangle$ is the content of the first $3$ cells of the tape and the last quantum state $|0\rangle$ represents the position of the tape head. It can be seen from the QSTD in Fig.\ref{fig4} that the SR-QTM has $4$ steps for all branches of computation. The transform process of the configuration is represented as follows.
\begin{eqnarray*}
&&|q_0\rangle|\#\rangle(a|00\rangle+b|01\rangle+c|10\rangle+d|11\rangle)|0\rangle\\
&\stackrel{1}{\longrightarrow}&|q_1\rangle|\#\rangle(a|00\rangle+b|01\rangle+c|10\rangle+d|11\rangle)|1\rangle\\
&\stackrel{2}{\longrightarrow}&|q_2\rangle|\#\rangle(a|00\rangle+b|01\rangle)|2\rangle+|q_3\rangle|\#\rangle(c|10\rangle+d|11\rangle)|2\rangle\\
&\stackrel{3}{\longrightarrow}&|q_4\rangle|\#\rangle(a|00\rangle+b|01\rangle)|1\rangle+|q_5\rangle|\#\rangle(c|11\rangle+d|10\rangle)|1\rangle\\
&\stackrel{4}{\longrightarrow}&|q_f\rangle|\#\rangle(a|00\rangle+b|01\rangle+c|11\rangle+d|10\rangle)|0\rangle.
\end{eqnarray*}

After the $4$ steps, the final configuration is achieved. From the final configuration, it is obvious that the QTM completes a CNOT operation between cell $1$ and cell $2$, where cell $1$ acts as control qubit and cell $2$ acts as target qubit.

\begin{figure}[htb!]
\begin{center}
\includegraphics[scale=0.95]{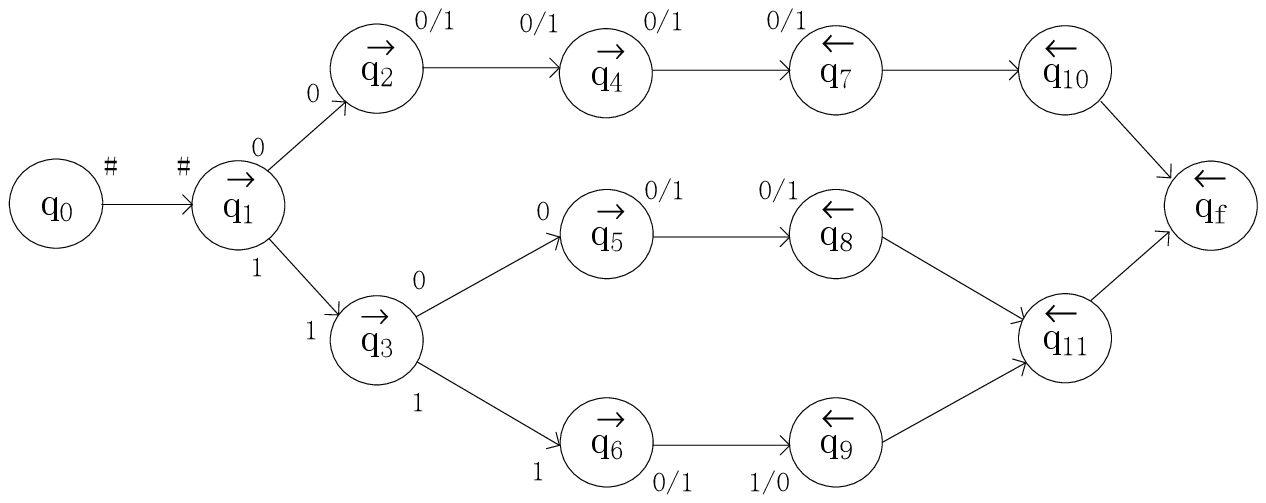}
\end{center}
\vspace{-3mm}
\caption{\label{fig5}The QSTD of SR-QTM which carries out Toffoli operation among cell $1$, cell $2$ and cell $3$. Cell $1$ and cell $2$ act as control qubit, and cell $3$ acts as target qubit.}
\end{figure}

Similar to the above construction, the SR-QTM simulating Toffoli gate can be
constructed and its QSTD is shown in Fig.\ref{fig5}. The set of states is $Q=\{q_0, q_1, q_2, q_3, q_4, q_5, q_6, q_7, q_8, q_9, q_{10}, q_{11}, q_f\}$. Suppose cell $1$ and cell $2$ of the tape act as control qubit, and cell $3$ acts as target qubit.

This way to construct SR-QTM for simulating Toffoli gate can be extended to any generalized Toffoli gate, which has more than $2$ control qubit.

Similar to the construction in Fig.\ref{fig3}, we can also construct SR-QTM which carries out CNOT operation between cell $i$ and cell $j$ or Toffoli operation among cell $i$, cell $j$ and cell $k$, where $i,j,k$ are arbitrary positive integers and each one is not equal to another.

{\bf Theorem 1}. For any quantum circuit consists of CNOT and single-qubit gates, there exists a SR-QTM which can efficiently simulate this quantum circuit.

Proof. Any single-qubit unitary transformation can be decomposed into single-qubit rotations $R_y(\theta)$ and $R_z(\theta)$ \cite{barenco95,nielsen00}. The single-qubit rotation $R_y(\theta)$ or $R_z(\theta)$ can be implement with quantum circuit consists of CNOT gate and other gates in $\{R_z(\pm\frac{\pi}{2^j}), R_y(\pm\frac{\pi}{2^j}), j\in{\mathbb N}\}$ within any precision \cite{liang11}. Thus, given any precision, arbitrary quantum circuit consists of CNOT and single-qubit gates can be decomposed into CNOT gate and other gates in $\{R_z(\pm\frac{\pi}{2^j}), R_y(\pm\frac{\pi}{2^j}), j\in{\mathbb N}\}$.

Then the following 3 steps are taken to obtain a SR-QTM which can efficiently simulate the quantum circuit.

(1) The gates in the quantum circuit are numbered with $0,1,2,\ldots$ in orders from the left to the right and from the top to the bottom of the circuit.

(2) Constructing a SR-QTM for each gate which perform either single-qubit rotation on cell $i$, or CNOT operation on cell $j$ and $k$.

(3) Dovetailing these SR-QTMs one by one in order $0,1,2,\ldots$.

It is obvious that the obtained QTM is also SR-QTM and can efficiently simulate the quantum circuit. In addition, each SR-QTM simulating each quantum gate has polynomial states and takes polynomial running time. Therefore, there exists a SR-QTM which can efficiently simulate any quantum circuit consisting of CNOT and single-qubit gates.

\section{Universal implementation of near-trivial transformation with QTM}
\label{}
According to Bernstein and Vazirani \cite{bernstein97}, near-trivial transformation is defined as follows.

{\bf Definition~4}. A unitary matrix $M$ is near-trivial if it satisfies one of the following two conditions.
\begin{enumerate}
 \item $M$ is the identity except that one of its diagonal entries is $e^{i\theta}$ for some $\theta\in[0,2\pi]$. For example, $\exists j,M_{jj}=e^{i\theta}$,$\forall k\neq j,M_{kk}=1$, and $\forall k\neq l,M_{kl}=0$.
 \item $M$ is the identity except that the submatrix in one pair of distinct dimensions $j$ and $k$ is the rotation by some angle $\theta\in[0,2\pi]$:$\left(\begin{array}{cc}cos\theta & -sin\theta \\ sin\theta & cos\theta\end{array}\right)$. So, as a transformation $M$ is near-trivial if there exists $\theta$ and $i\neq j$ such that $Me_i=(cos\theta)e_i+(sin\theta)e_j$,$Me_j=(-sin\theta)e_i+(cos\theta)e_j$, and $\forall k\neq i,j,~Me_k=e_k$.
\end{enumerate}

\begin{figure}[htb!]
\begin{center}
\includegraphics[scale=0.65]{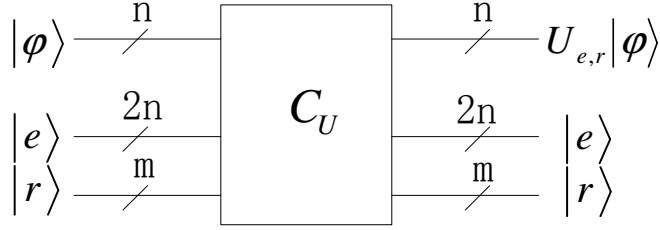}
\end{center}
\vspace{-4mm}
\caption{\label{fig7}Universal quantum circuit implementing near-trivial transformation. This circuit consists of only CNOT, generalized Toffoli and some single-qubit gates in $\{R_z(\pm\frac{\pi}{2^j}), R_y(\pm\frac{\pi}{2^j}), j\in{\mathbb N}\}$. $|\varphi\rangle$ is any $n$-qubit data. The input $|e\rangle\otimes|r\rangle$ is the encoding of the near-trivial transformation $U_{e,r}$. The quantum register $r$ is a $m$-qubit register, and the value of $m$ is determined by the precision of near-trivial transformation.}
\end{figure}

In our paper \cite{liang11}, a universal quantum circuit is constructed to implement any near-trivial transformation, where its universality is in the meaning of Bera et al.\cite{bera08}. The quantum circuit can be shown simply in Fig.\ref{fig7}. This universal quantum circuit consists of some CNOTs, generalized Toffoli and single-qubit gates in $\{R_z(\pm\frac{\pi}{2^j}), R_y(\pm\frac{\pi}{2^j}), j\in{\mathbb N}\}$. The input of the circuit consists of two parts. The first part is a register for inputting quantum data such as $|\varphi\rangle$. The second part includes two registers for inputting an encoding $|e\rangle\otimes|r\rangle$ of arbitrary near-trivial transformation, where $|e\rangle$ is the encoding of the dimensions, and $|r\rangle$ is the encoding of the angle. This circuit can implement any near-trivial transformation with any accuracy, and the accuracy is determined by the length of the encoding $|r\rangle$.

Here, we can present a universal implementation with SR-QTM according to the result in Section \ref{section3}.
In Fig.\ref{fig7}, the quantum circuit consists of only of only CNOT, generalized Toffoli and some single-qubit gates in $\{R_z(\pm\frac{\pi}{2^j}), R_y(\pm\frac{\pi}{2^j})$, $j\in{\mathbb N}\}$. From the constructions in Section \ref{section3}, CNOT operation between $i$th qubit and $j$th qubit ($i\neq j$) and generalized Toffoli operation among more than $2$ qubits can be simulated by some SR-QTMs. We can also construct SR-QTM to simulate single-qubit gate, which performs $U\in\{R_z(\pm\frac{\pi}{2^j}), R_y(\pm\frac{\pi}{2^j})$, $j\in{\mathbb N}\}$ on the $i$th qubit. Thus, by dovetailing these SR-QTMs in order, a big SR-QTM is obtained and it can efficiently simulate the universal quantum circuit in Fig.\ref{fig7}.

\section{Accepting languages with SR-QTM}
\label{}
{\bf BQP} \cite{bernstein97} is a class of languages that is decidable with bounded error on some polynomial-time QTM. We define the class {\bf SR-BP} as the set of languages which are decided by polynomial-time SR-QTM with bounded error.
{\bf BUPQC} is a class of languages that are decidable by a family of polynomial-size uniform quantum circuit, and {\bf BUPQC}={\bf BQP} \cite{nishimura02}. Let $L$ be a language in {\bf BUPQC}.

{\bf Theorem 2}. {\bf SR-BP}={\bf BQP}.

{\bf Proof.} By the definition of SR-QTM, SR-QTM is a special case of QTM. So, {\bf SR-BP}$\subseteq${\bf BQP}. According to \cite{nishimura02}, it holds that {\bf BQP}={\bf BUPQC}. Moreover, it can be concluded from Theorem 1 that {\bf BUPQC}$\subseteq$ {\bf SR-BP}. Therefore, {\bf SR-BP}={\bf BUPQC}={\bf BQP}.$\hfill{}\Box$

This theorem establishes the computational equivalence (in the bounded error setting) between QTM and SR-QTM, which is a simple kind of QTM. Since SR-QTM halts deterministically, the halting scheme problem dissolved in this sense.

As an example of our result, we consider the synchronization theorem proved by Bernstein and Vazirani \cite{bernstein97}. The synchronization theorem proposed a class of functions which is a subset of {\bf P} and can be efficiently computed by stationary normal form QTM. Moreover, from the Theorem 2 and {\bf P}$\subseteq$ {\bf BQP}, it can be deduced that {\bf P}$\subseteq$ {\bf SR-BP}. Thus, this class of functions can be efficiently computed by SR-QTM, which is a special case of stationary normal form QTM. The result is given in this corollary.

{\bf Corollary 1}. If $f$ is a function mapping strings to strings which can be computed in deterministic polynomial-time and such that the length of $f(x)$ depends only on the length of $x$, then there is a SR-QTM $M$ which given input $x$, produces output $x; f(x)$, and whose running time depends only on the length of $x$.

\section{Discussion}
\label{}
We have constructed a universal quantum circuit to implement near-trivial transformation in another paper \cite{liang11}. Since SR-QTM can efficiently simulate any quantum circuit consisting of CNOT, generalized Toffoli and single-qubit gates, we can present a universal implementation of near-trivial transformation with SR-QTM. This means that there exists a QTM (exactly saying SR-QTM) which is universal for a subclass of QTM (those QTMs implementing near-trivial transformations). Moreover, any unitary transformation can be decomposed into a product of several near-trivial transformations \cite{bernstein97}, so SR-QTM can realize any unitary transformation.

An important feature of SR-QTM is that it has the same time steps for all different branches of computation. Thus, the halting of SR-QTM is not probabilistic but deterministic, and the halting scheme problem proposed by Myers \cite{myers97} does not exists for SR-QTM.

The computational power of SR-QTM is considered and it is proved the class {\bf SR-BP} of languages that are decidable by polynomial SR-QTM is equal to {\bf BQP}. This establishes the computational equivalence (in the sense of exactly deciding) between ordinary QTM and SR-QTM. It means that any language which can be exactly decidable with a ordinary polynomial-time QTM can also be exactly decidable with a polynomial-time QTM which halts deterministically and has deterministic tape head position.

\section{Conclusions}
We introduce a subclass of QTM named SR-QTM and propose a new tool named QSTD to intuitively describe SR-QTM. QSTD is helpful in further understanding QTM. Because SR-QTM halts deterministically, there does not exist halting scheme problem for SR-QTM. Then we show how to construct SR-QTM to simulate quantum circuit. Moreover, based on universal quantum circuit of near-trivial transformations, we construct a SR-QTM which is universal for all near-trivial transformations. This means that there exists a QTM which is universal for a subclass of QTM. Finally, we define {\bf SR-BP} as the class of languages that are decidable by polynomial-time SR-QTM, and prove that {\bf SR-BP}={\bf BQP}, so SR-QTM is computational equivalent with ordinary QTM in the bounded error setting.

\section*{Acknowledgement}
This work was supported by the National Natural Science Foundation of China under Grant No.61173157.






\end{document}